\documentclass[aps,pra,twocolumn,superscriptaddress,floatfix]{revtex4}
\usepackage{graphicx, amssymb}

\begin{document}

\title{A lattice of double wells for manipulating pairs of cold atoms}

\author{J. Sebby-Strabley}
\affiliation{National Institute of Standards and Technology,
       Gaithersburg, Maryland 20899, USA }
\author{M. Anderlini }
\affiliation{National Institute of Standards and Technology,
       Gaithersburg, Maryland 20899, USA }
\author{P. S. Jessen}
\affiliation{Optical Sciences Center, University of Arizona, Tucson,
Arizona 85721, USA}
\author{J. V. Porto}
\affiliation{National Institute of Standards and Technology,
       Gaithersburg, Maryland 20899, USA }

\begin{abstract}
We describe the design and implementation of a $2$D optical lattice
of double wells suitable for isolating and manipulating an array of
individual pairs of atoms in an optical lattice.   Atoms in the square
lattice can be placed in a double well with any of their four nearest
neighbors.  The properties of the double well (the barrier height and
relative energy offset of the paired sites) can be dynamically controlled.
The topology of the lattice is phase stable against phase noise imparted by
vibrational noise on mirrors.  We demonstrate the dynamic control of the
lattice by showing the coherent splitting of atoms from single wells into
double wells and observing the resulting double-slit atom diffraction pattern.
This lattice can be used to test controlled neutral atom motion among lattice sites
and should allow for testing controlled two-qubit gates.
\end{abstract}

\maketitle

Bose Einstein condensates (BEC) in optical lattices have proven to
be an exciting and rich environment for studying many areas of
physics, such as condensed matter physics, atomic physics, and
quantum information processing (see for instance
\cite{Bloch2005}). Optical lattices are very versatile because
they allow dynamic control of many important experimental
parameters. Dynamic control of the amplitude of the lattice has
been widely used (e.g.
\cite{Denschlag2002,Schori2004,Greiner2002,Peil2003}); recent
experiments have used a state dependent lattice to dynamically
control the geometry and transport of atoms in the lattice
\cite{Mandel2003}. Recently there have been several proposals for
using optical lattices to perform neutral atom quantum computation
\cite{Brennen1999,Jaksch1999,Calarco2004}. With optical lattices
it should be possible to load single atoms into individual lattice
sites with high fidelity \cite{Pupillo2004}, and then to isolate
and manipulate pairs of atoms confined by the lattice in order to
perform 2-qubit gates. Loading of single atoms into lattice sites
or traps was demonstrated by
\cite{Greiner2002,Alt2003,Darquie2005,Albiez2005,Schumm2005}, but
to date no neutral atom based trap can isolate and control
interactions between individual pairs of atoms.  While previous
experiments have demonstrated the clustered entanglement of many
atoms confined by an optical lattice \cite{Mandel2003b}, the
unique ability to isolate and control interactions between pairs
of atoms would allow for entanglement between just the pair of
atoms.

In this paper we report on a double well optical lattice designed to
isolate and control pairs of atoms. The lattice is constructed from
two 2D lattices with different spatial periods, resulting in a 2D
lattice whose unit cell contains two sites.  Within the pair, the
barrier height and relative depths of the two sites are
controllable. Furthermore, the orientation of the unit cell can be
changed, allowing each lattice site to be paired with any one if its
four nearest neighbors. The double well lattice is phase stable in
that its topology is not sensitive to phase noise from motion of the
mirrors.  This lattice, in combination with an independent 1D
lattice in the third direction to provide 3D confinement, is ideal
for testing many 2 qubit ideas, particularly quantum computation
based on the concept of ``marker atoms'' \cite{Calarco2004} and
controlled collisions \cite{Jaksch1999}. Among other applications,
this lattice could be used for studying tunnel coupled pairs of 1D
systems, interesting extensions to the Bose Hubbard model
\cite{Scarola2005}, and quantum cellular automata
\cite{Brennen2003}.

This paper is divided into six sections.  In Section I we discuss
the ideal structure of the lattice.  Section II describes several
experimental issues which need to be considered in order to
experimentally realize an ideal double well lattice. Section III
details the experimental realization of this lattice and a
measurement of the important parameters. In Section IV we show the
momentum components present in our lattice by mapping the lattice
Bruillion zone. In section V we demonstrate the dynamic control of
the properties and topology of the double well lattice by showing
the coherent splitting of atoms from a single well into a double
well. We summarize and present prospective applications in Section
VI.

\section{Idealized 2D double-well lattice}

\begin{figure}
\includegraphics[scale=.75]{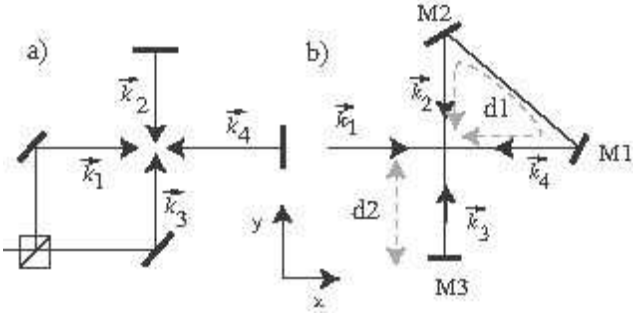}
\caption{2D lattices with four beams.(a) Lattices formed by
interfering two independent standing waves must be actively
stabilized to be topologically phase stable against phase noise
caused by vibration of mirrors.  (b) Lattices formed from a folded
retroreflected beam have intrinsic topological phase stability. }
\label{simple}
\end{figure}

An ideal double-well lattice would allow for atoms in neighboring
pairs of sites to be brought together into the same site,
requiring topological control of the lattice structure.  It has
been shown \cite{Grynberg1993} that a D-dimensional optical
lattice created with no more than D+1 independent light beams is
topologically stable to arbitrary changes of the relative phases
of the D+1 beams.  This geometry is usually preferred since phase
noise (e.g. that imparted by vibrational noise on mirrors) will
merely cause a global translation of the interference pattern. To
allow for topological control, a general double-well lattice will
necessarily have more than D+1 beams, but it would be desirable to
preserve the topological insensitivity due to mirror-induced phase
noise. To achieve vibrational phase stability in a D-dimensional
lattice made with more than D+1 beams, one can actively stabilize
the relative time phase between standing waves \cite{Greiner2001}
\cite{Hemmerich1992}. Alternatively the lattice can be constructed
from a folded, retroreflected standing wave, which forces the
relative time phase between standing waves to be a constant
~\cite{Rauschenbeutel1998}. Examples for a 2D case are shown in
Fig. \ref{simple}.

In this paper we consider the latter design, shown in Fig.
\ref{simple} b.  In this scheme, the same laser beam intersects
the position of the atom cloud four times.  The incoming beam with
wave vector $\vec{k}_1$ along $\hat{x}$ is reflected by mirrors M1
and M2, and after traveling an effective distance $d_1$ (where the
effective distance includes possible phase shifts from the
mirrors) returns to the cloud with wave vector $\vec{k}_2$. The
beam is then retroreflected by M3, returning a third time with
wave vector $\vec{k}_3 = -\vec{k}_2$, having traveled an
additional effective distance $2 d_2$. Finally, it makes a fourth
passage with $\vec{k}_4 = -\vec{k}_1$, traveling again the
distance $d_1$. The total electric field for this 2D 4-beam lattice is given by Re[$\vec{E}(x,y) e^{i \omega t}$], where
\begin{eqnarray}
&&\vec{E}(x,y) = E_1 e^{i \vec{k}_1 \cdot \vec{r}} \hat{e}_1 + E_2
e^{i(\theta + \vec{k}_2 \cdot \vec{r})} \hat{e}_2\nonumber
\\   && \ \ \ \ + E_3 e^{i(\vec{k}_3 \cdot \vec{r} + \theta + 2 \phi)} \hat{e}_3 + E_4 e^{i(2
\theta + 2 \phi + \vec{k}_4 \cdot \vec{r} )} \hat{e}_4,
 \label{egeneral}
\end{eqnarray}
and $\vec{r} = x \hat{x} + y \hat{y}$, $\theta = k d_1$, $\phi = k
d_2$, $k = 2 \pi / \lambda$ ($\lambda$ is the wavelength of the
lattice light), and $\hat{e}_{i}$ is the polarization vector of
the $i^{th}$ beam. In the absence of polarization rotating
elements and ignoring polarization dependent phase shifts from
mirrors, $\hat{e}_4 = \hat{e}_1$ and $\hat{e}_3 = \hat{e}_2$.
Since the beam retraces the same path, there are only two
independent relative phases between the four beams.  As a result,
the lattice is topologically stable to vibrational motion of M1,
M2, and M3; variations in $d_1$ and $d_2$ result in a simple
translation of the interference pattern~\cite{Rauschenbeutel1998}.

The potential seen by an atom in a field Re[${\vec E} e^{i\omega
t}$] is given by $U = - (1/4) \vec{E}^* \cdot \mbox{\boldmath $
\alpha$} \cdot \vec{E}$, where \mbox{\boldmath $ \alpha$} is the
atomic polarizability tensor~\cite{Deutsch1998}. In general,
\mbox{\boldmath $ \alpha $} depends on the internal (angular
momentum) state of the atom, having irreducible scalar, vector, and
2$^{nd}$ rank tensor contributions with magnitudes $\alpha_s$,
$\alpha_v$ and $\alpha_t$, respectively. The scalar light shift,
$U_s = -\alpha_s |\vec{E}|^2/4$, is state independent and directly
proportional to the total intensity. The vector light shift, $U_v =
i \alpha_v (\vec{E}^* \times \vec{E})\cdot \hat{F}/4$, depends on
the projection of total angular momentum $\hbar \hat{F}$. It can be
viewed as arising from an effective magnetic field whose magnitude
and direction depend on the local ellipticity of the laser
polarization, $\vec{B}_{\rm eff} \sim i \alpha_v (\vec{E}^* \times
\vec{E})$. It vanishes for linearly polarized light. The total
vector shift in the presence of a static magnetic field $\vec{B}$ is
determined from the energy of an atom in the vector sum field
$\vec{B}_{\rm eff}+\vec{B}$. The 2$^{nd}$-rank tensor contribution
is negligible for ground state alkali atoms far detuned with respect
to hyperfine splittings \cite{Deutsch1998}, and we will ignore it in
this paper.

\begin{figure}
\includegraphics[scale=.55]{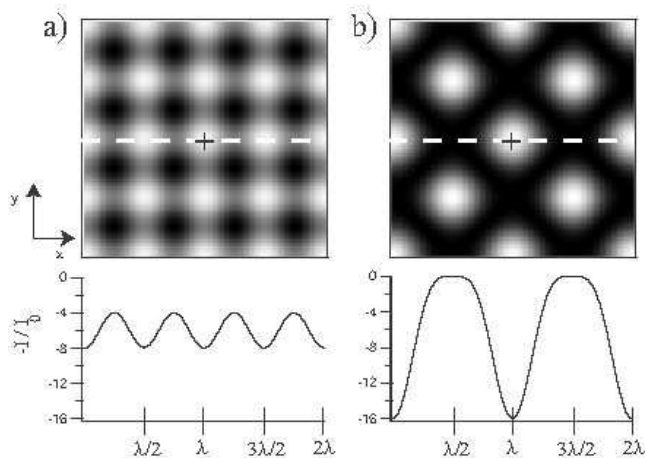}
\caption{Calculated intensities for in-plane lattice (a) and the
out-of-plane lattice (b).  Cross sections taken on the white dashed
line are shown below their respective plot; a cross is used to
denote the origin in each plot.  The in-plane lattice has the
familiar cos$^2$ profile typical of $\lambda/2$ lattices, while the
out-of-plane lattice has a cos$^4$ profile and periodicity of
$\lambda$.  The flat portion of the (b) cross section shows the
intersection of two nodal lines. } \label{lattices}
\end{figure}

Consider the ideal situation with four beams of equal intensities ($E_i
= E$) which intersect orthogonally ($\vec{k}_1 \cdot \vec{k}_2
= 0$).  As a first case consider $\hat{e}_1 = \hat{y}, \hat{e}_2 =
\hat{x}$, where all the light polarizations are in-the-plane.  We
will refer to this configuration as the ``in-plane'' lattice. The spatial dependence of the electric field is given by the real part of
\begin{eqnarray} \vec{E}_{xy}(x,y)  &=& E \left(  e^{i k x} +
    e^{i(2 \theta_{xy} + 2 \phi_{xy} - k x)} \right)\hat{y} \nonumber \\
   && \ \ \ \ \   +
E \left(e^{i (-k y+\theta_{xy})} +
      e^{i(\theta_{xy} + 2 \phi_{xy} + k y)} \right)
     \hat{x},\label{efieldd} \nonumber
\end{eqnarray}
where $\theta_{xy}$ and $\phi_{xy}$ are the path length
differences for in-plane light taking into account that the path
length difference could be polarization dependent.  This gives a
normalized total intensity of
\begin{eqnarray}
I_{xy}(x,y)/I_0  &=& 2 \cos{(2 k x - 2 \theta_{xy}- 2 \phi_{xy})} \nonumber\\
&& \ \ \ \ \ + 2  \cos{(2 k y + 2 \phi_{xy}) + 4 }\label{Ixy}
\end{eqnarray}
where $I_0$ is the intensity of a single beam. Due to the
orthogonal intersection ($\vec{k}_1 \cdot \vec{k}_2 = 0$, etc.)
and the orthogonality of the polarizations between $\vec{k_1}$ and
$\vec{k_2}$ etc., the resulting four beam lattice is the sum of
two independent 1D lattices.  As shown in Fig. \ref{lattices}a,
this creates a 2D square lattice with anti-nodes (and nodes)
spaced by $\lambda$/2 along $\hat{x}$ and along $\hat{y}$. Since
the four beam intensities are equal, the lattice forms a perfect
standing wave, and the polarization is everywhere linear, although
the local axis of linear polarization changes throughout the
lattice. In this case the vector light shift vanishes, and the
light shift is strictly scalar $U(x,y) = - \alpha_s \epsilon_0
|E(x,y)|^2/4$. Note from Eq. \ref{Ixy} that varying $\theta_{xy}$
changes the relative position of the lattice formed by $\vec{k}_1$
and $\vec{k}_4$, moving the lattice along $\hat{x}$. The phase
$\phi_{xy}$ affects both 1D lattices, shifting the combined 2D
lattice along $(\hat{x} - \hat{y})/\sqrt{2}$.

As a second case consider ($\hat{e}_1 = \hat{e}_2 =\hat{z}$),where
all the light polarizations are out-of-the-plane.  We will refer
to this configuration as the ``out-of-plane'' lattice.   The
electric field is given by the real part of
\begin{eqnarray} \vec{E}_z(x,y) & = &  E( e^{i k x} +
e^{i(2 \theta_{z} + 2 \phi_{z} - k x)}  \nonumber\\ && \ \ \ \ \ +
 e^{i (-k y+\theta_{z})}  +
     e^{i(\theta_{z} + 2 \phi_{z} + k y)} ) \hat{z}.\nonumber
\end{eqnarray}
where $\theta_{z}$ and $\phi_{z}$ are the path length differences
for out-of-plane light. In this case the intensity is not simply a
sum of independent functions of x and y, but rather given by
\begin{eqnarray}
I_z(x,y)/I_0 & = & 4 \left[\cos{(k x- \theta_z -\phi_z)} + \cos{(k y + \phi_z)}\right]^2 \nonumber \\
& = & 16\left[ \cos\left(\frac{k}{2}( x + y)-\frac{\theta_z}{2}
\right)\right]^2\ \nonumber \\ && \ \ \ \ \times \left[\cos
\left(\frac{k}{2} (x - y)-\frac{\theta_z}{2} -  \phi_z
\right)\right]^2. \label{Iz}
\end{eqnarray}
As shown in Fig \ref{lattices}b, the added interference creates
components at $k$ in addition to the components at $2 k$ resulting
in a lattice spacing along $\hat{x}$ and $\hat{y}$ of $\lambda$
rather than $\lambda/2$ (the lattice period along $\hat{x} +
\hat{y}$ is $\lambda/\sqrt{2}$). In addition, the nodal structure
changes in that there are nodal lines along the diagonals. In
particular, every other anti-node of the in-plane lattice is at
the intersection of two nodal lines in the out-of-plane lattice.
The polarization is everywhere linear along $\hat{z}$, giving rise
to a strictly scalar light shift. As with the in-plane lattice,
varying $\theta_{z}$ translates the out-of-plane lattice along
$\hat{x}$, and varying $\phi_z$ translates the lattice along
$(\hat{x} - \hat{y})/\sqrt{2}$.

A double well lattice is realized by combining the in-plane and
out-of-plane polarizations.   Since the polarizations of the two
lattices are orthogonal, the total intensity is $I_{tot} =
I_{xy}+I_{z}$, and the scalar part of the light shift is simply a
sum of the light shifts from the in-plane and out-of-plane
lattices. Electro-optic elements in the beam paths $d_1$ and $d_2$
can produce different phase shifts for different input
polarization, allowing for control of the relative phases $\delta
\theta = \theta_z - \theta_{xy}$ and $\delta \phi = \phi_z -
\phi_{xy}$, while maintaining vibrational phase stability of the
combined lattice. This combined lattice can have a vector light
shift, since relative phase shifts between the two polarizations
allow for non-zero ellipticity, $i(\vec{E}^* \times \vec{E}) \neq
0$. If both lattices are everywhere in time-phase ($\delta \theta
= 0$ or $\pi$ and $\delta \phi = 0$ or $\pi$), the vector shift
vanishes. Otherwise, there is a non-zero, position dependent
$\vec{B}_{\rm eff}(x,y)$ which lies in the $\hat{x}$-$\hat{y}$
plane.

\begin{figure}
\includegraphics[scale=.65]{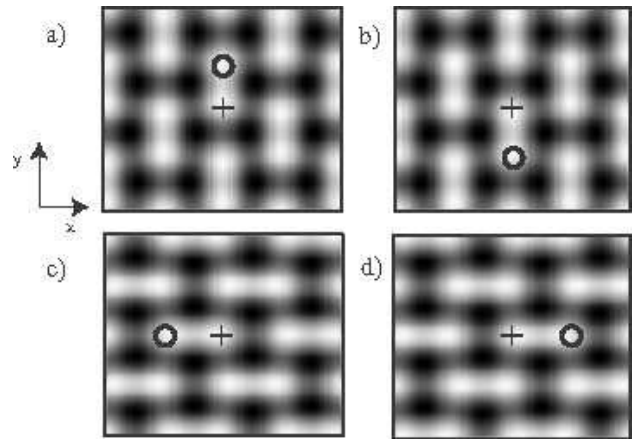}
\caption{ Adjustment of the phases $\delta \theta$ and $\delta
\phi$ allow for nearest neighbor pairing with all four nearest
neighbors. ``+'' marks the location of a lattice site located at
the origin which can be paired with any of its four nearest
neighbors (shown with $\bigcirc$) depending upon the choice of
phase: a) $\delta\theta = \pi/2,\delta \phi = -\pi/2$ b)
$\delta\theta = -\pi/2, \delta\phi = \pi/2$ c) $\delta\theta =
-\pi/2, \delta\phi = 0$ d) $\delta\theta = \pi/2, \delta\phi = 0$.
}\label{neighbors}
\end{figure}

\begin{figure}
\includegraphics[scale=.6]{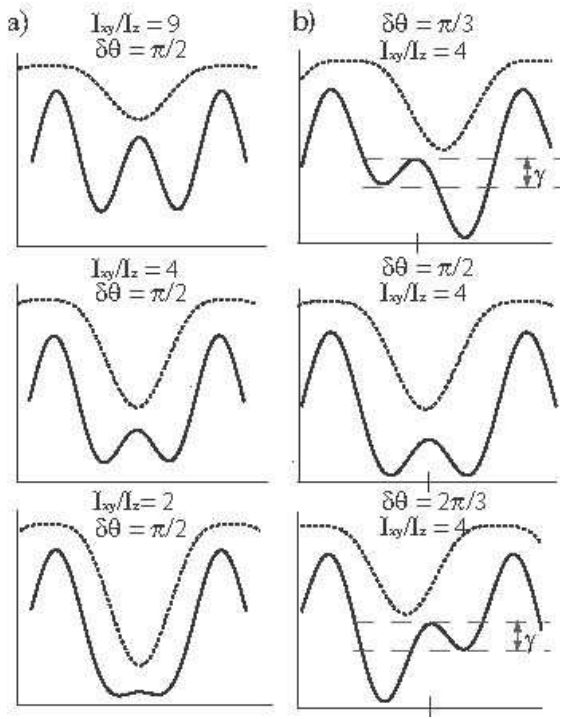}
\caption{ Cross sections of example double well potentials.  Solid
line represents the double well potential; dotted line shows the
placement and amplitude of the out-of-plane lattice. a) The
barrier height, labeled above by the quantity $\gamma$, of the
double well can be adjusted by placing the out-of-plane lattice
``in the barrier'' and adjusting the ratio of $I_{xy}/I_{z}$. b)
The ``tilt'' of the double well (the relative offset between
adjacent sites) can be changed by adjusting $\delta\theta$ and
$\delta\phi$.}\label{tilt}
\end{figure}

Control of the phase shifts, $\delta\phi$ and
$\delta\theta$, and the relative intensity, $I_{xy}/I_{z}$,
provides the flexibility to adjust the double-well parameters:
the orientation (which wells are paired), the barrier height, and
the tilt. For instance, double-well potentials along the
$\hat{x}$-direction can be formed by setting $\delta \phi=0$ and
$\delta \theta=\pi/2$. Fig. \ref{neighbors} demonstrates how a
site can be paired with any one of its four nearest neighbors.
Control of the barrier height and of the tilt are shown in Fig.
\ref{tilt}.

\section{Realistic 2D Double Well Lattice }

In the previous section we considered idealized lattices, making assumptions about the amplitudes,
wave-vectors and polarizations of the beams in the lattice.  In this section we discuss considerations
needed to experimentally realize the lattices described above.

\subsection{In-plane lattice}
For certain applications, such as the realization of the Mott-insulator state \cite{Greiner2002}, we need a nearly perfect
in-plane lattice, namely a square 2D lattice with little or no energy
offsets between neighboring sites. There are three primary sources
of imperfections that affect the performance of the in-plane
lattice: imperfect control of the input polarization ($\hat{e_i}
\cdot \hat{z} = \sin\beta \neq 0$), imperfect alignment causing
the beams to be nonorthogonal, ($\vec{k}_1 \cdot \vec{k}_2 = \sin
\epsilon \neq 0$), and imperfect intensity balance among all four
beams ($E_1 \neq E_2 \neq E_3 \neq E_4$).

\begin{figure}
\includegraphics[scale=.4]{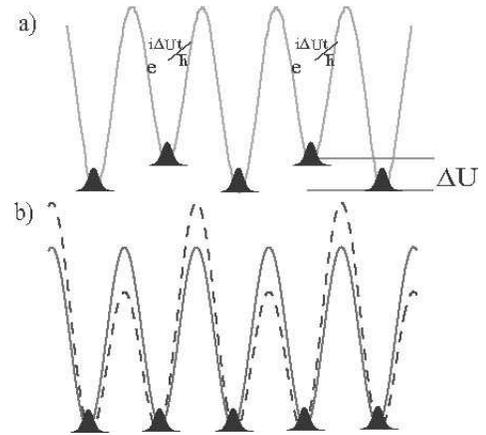}
\caption{  Lattice imperfections causing  a) modulation of the
lattice depth, $\Delta U$, between neighboring sites and b) state
dependent modulation of the barrier height by a polarization
lattice. In b) the solid line is the cross section of the intensity
lattice; the dashed line is the cross section of the state dependent
lattice resulting from unbalanced beam intensities. Atoms in the
ground state of each well are shown schematically.
}\label{imperfections}
\end{figure}

When trying to make a perfect in-plane lattice, if the input
polarization is tilted by an angle $\beta$ with respect to the
$xy$-plane, then there is a $\hat{z}$ component to the light. The
result is a contamination of the in-plane lattice by an out-of-plane
lattice that modulates the lattice depth with a periodicity of
$\lambda$ (Fig. \ref{imperfections}a). Neighboring sites will
experience an energy shift $\Delta U = 4 U_0
\sin^2\left(\beta\right)$ where $U_0$ is the depth of a $\beta = 0$
in-plane lattice.  Since $\Delta U$ scales as $\beta^2$ for small
$\beta$, the in-plane lattice is fairly tolerant to small rotations
of the input polarization.  For example, a misalignment of 10 mrad
will cause a 0.04\% modulation of the trap depth.

The more stringent demand for minimizing site-to-site offsets of the
in-plane lattice is the orthogonality of the two standing waves. If
$\vec{k}_1 \cdot \vec{k}_2 = \sin\epsilon \neq 0$, standing waves
$\vec{k}_1,\vec{k}_4$ and $\vec{k}_2,\vec{k}_3$ have nonorthogonal
polarization and give rise to an interference term in the total
intensity, thus causing an energy offset between neighboring sites
given by $\Delta U \approx 4 U_0 \epsilon $ for small $\epsilon$
(Fig. \ref{imperfections}a). This imperfection has the same effect
as imperfect input polarization, but is harder to minimize since it
scales linearly with $\epsilon$.  For example, a misalignment of 10
mrad will cause a 4\% modulation of the trap depth. We describe
below how to control both imperfections.

The third source of imperfections for the in-plane lattice is the
intensity imbalance between the four beams. Experimentally,
intensity imbalance can arise from reflection and transmission
losses along the beam path as well as from unequal beam waists at
the intersection \footnote{Intensity imbalances can be alleviated
by focusing the beam to a smaller beam waist with each passage
through the atom cloud so that the intensity at the atom cloud is
held the same.}. In general, light imbalance breaks the symmetry
between the $x$ and $y$ direction, which removes the degeneracy
between the vibrational excitations along $x$ and $y$. Typically,
this does not adversely affect the lattice. We also note that
since the beam experiences the same losses while traversing $d_1$
each time, then for equal beam waists $E_1 E_4=E_2 E_3$, and the losses do not produce well asymmetries.

A more important consequence of intensity imbalance is that the total field is not everywhere linearly polarized, but
rather has some ellipticity,
\begin{eqnarray}
& & \frac{i}{2}(\vec{E}^* \times \vec{E})=
 \left[(E_1 E_2 - E_3 E_4) \sin ( k x + k y - \theta  ) \right. \nonumber \\
& & \ \ \ \ \ + \left. \left( E_1 E_3 - E_2 E_4 \right) \sin (k x
- ky - \theta  - 2 \phi )\right] \hat{z}.
     \label{pol_diff}
\end{eqnarray}
This causes a state dependent spatially varying vector light shift
with period $\lambda$, even in the absence of the out-of-plane
lattice. As evident from Eq. \ref{pol_diff}, for perfect intensity
balance the ellipticity will vanish, resulting in purely linear
polarization. Comparing Eq. \ref{Ixy} with Eq. \ref{pol_diff} one
can see that the phase of the polarization lattice is spatially
out of phase with the intensity lattice (see Fig.
\ref{imperfections}b) resulting in a state dependent barrier
height between lattice sites, with relatively little modification
of the potential near the minima.

\subsection{Out-of-plane lattice}
In general, the structure of the out-of-plane lattice is fairly
robust against the three imperfections mentioned above.  A minor
consequence of field imbalance is the possible disappearance of
perfect nodal lines. One finds, for example, that at the position
of the nodal line intersection, the intensity becomes
\begin{equation}
I_{z, min} = \frac{c \epsilon_0}{2} \left(E_1 - E_2 - E_3 + E_4
\right)^2
\end{equation}
where $\epsilon_0$ is the electric constant (permittivity of free
space), and $c$ is the speed of light in vacuum. There are exact
nodes when $E_1+E_4 = E_2+E_3$, and this condition is trivially
satisfied when the light fields are balanced.  As with the
in-plane lattice, degeneracies of vibrational excitations are
lifted when the intensities are imbalanced.

\subsection{The double-well lattice}
A composite in-plane and out-of-plane lattice can be made by
adjusting the angle $\beta$ to control the admixture of the two
components. For the combined lattice, the consequences and control
of imperfections are similar to the in-plane lattice. With the added
flexibility to control intensity and relative phase, we can in fact
use $\beta$ and $\delta \theta$ to compensate for $\epsilon \neq 0$
(at least for a given magnetic sub-level).  The vector light shift
for an intensity imbalanced double-well lattice is somewhat more
complicated (yet easily calculable), having position dependent
ellipticity along $\hat{x}$, $\hat{y}$ and $\hat{z}$.  Many
experiments are carried out in the presence of a spatially uniform
bias field $\vec{B}$, so that the total field seen by the atoms is
given by the vector sum $\vec{B}_{\rm tot}= \vec{B}+ \vec{B}_{\rm
eff}$. For $| \vec{B}| >> | \vec{B}_{\rm eff}|$, the direction of
the quantization axis remains nearly constant along $\vec{B}$
throughout the lattice. The magnitude of the state dependent shift
in this limit is proportional to
\begin{eqnarray}
  |\vec{B}_{\rm  tot} | & = &  \sqrt{ \left(\vec{B}+\vec{B}_{\rm eff}\right)^2}
 \nonumber \\
 & \approx &
 | \vec{B} | + \vec{B}_{\rm eff} \cdot \left(\frac{\vec{B}}{| \vec{B} |}\right), \label{eq:Bfield}
\end{eqnarray}
and only the component of $\vec{B}_{\rm eff}$ along $\vec{B}$ contributes to the potential.
The ability to adjust the direction of $\vec{B}$ provides significant flexibility in designing
state-dependent potentials, and allows for state dependent motion of atoms between the two sites of the double-well.

\section{Implementation}

This double well lattice was implemented on an apparatus described
elsewhere \cite{Peil2003}.  $^{87}$Rb Bose Einstein condensates are
produced in an ultra high vacuum glass cell.  We use RF evaporation
to make BECs with $\approx 200,000$ atoms in the $F=1$, $m_F = -1$
hyperfine state. The BEC is confined in a cylindrically symmetric
magnetostatic trap with $\omega_{\perp}/2\pi = 24$ Hz and
$\omega_{\parallel}/2\pi = 8 $Hz. The Thomas-Fermi radii of
condensates are $\approx 15 \mu$m and $\approx 40 \mu$m
respectively, with mean-field atom-atom interaction energy
approximately 500 Hz. Atoms in the BEC are then directly loaded into
the ``tubes'' created by the 2D double well lattice potential.  The
lattice beams are derived from a continuous wave (CW) Ti:Sapphire
laser with $\lambda = 810$ nm, detuned far from the D1 (795 nm) and
D2 (780 nm) transitions in $^{87}$Rb. On average 2600 in-plane
lattice sites or 1300 out-of-plane lattice sites (tubes) are filled
with approximately 80 and 160 atoms per site respectively. Due to
the tight confinement, the mean-field energy is much larger in the
tubes than in the magnetic trap, as much as 7 kHz.  During our
experiments the magnetic confining potential is left on.

\begin{figure}
\includegraphics[scale=.6]{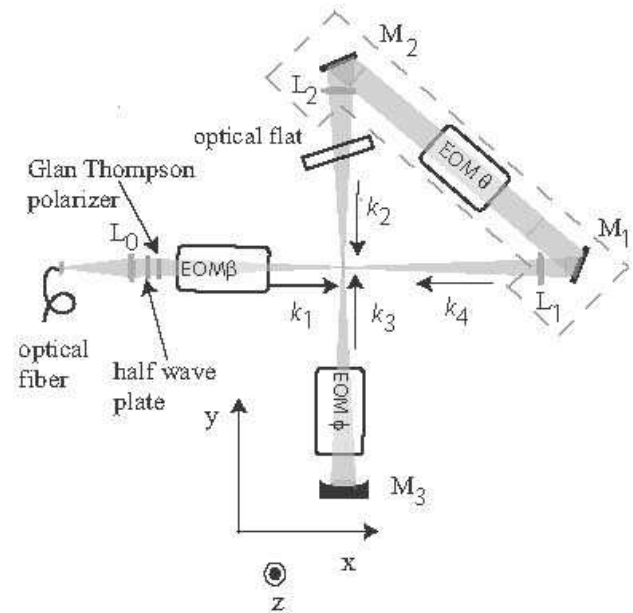}
\caption{  Schematic of the experimental implementation of the 2D
double well lattice made from a single folded, retroreflected beam.
Mirrors M1 and M2, lenses L1 and L2, and EOM$\theta$ are mounted on
a fixed plate.}\label{bowtie}
\end{figure}

The experimental schematic of the double well lattice is shown in
Fig. \ref{bowtie}. An acousto-optical modulator (AOM) provides rapid
intensity control of the lattice light. The lattice light is coupled
into a polarization maintaining fiber to provide a clean TEM$_{00}$
spatial mode.  A Glan-Thompson polarizer after the fiber creates a
well defined polarization in the $xy$-plane.  The light is folded by
plane mirrors M1 and M2 then retroreflected by concave mirror M3.
Lenses L$_0$, L$_1$, and L$_2$, in the input beam and after M1, M2
respectively provide a weak focus (all four beams have 1/e$^2$ beam
radius of $\approx 170$ $\mu$m) at the intersection of the four
beams. A 1 cm thick optical flat after L$_2$ is used to translate
the beam with wave vector $\vec{k}_2$ without changing the angle of
$\vec{k}_2$ relative to $\vec{k}_1$. Mirror M3 images the
intersection point back onto itself.

Three electro-optic modulators (EOMs):  EOM$\beta$, EOM$\theta$,
and EOM$\phi$, control the topology of the lattice. EOM$\beta$ is
aligned with its fast axis orientated 45$^{\circ}$ relative to the
axis of the Glan-Thompson polarizer, allowing for control of the
angle $\beta$, which determines the ratio $I_{xy}/I_z =
\cot^2\beta$. EOM$\theta$ and EOM$\phi$ are aligned with their
fast axes in the $xy$-plane allowing for control of the
differential phases $\delta\theta$ and $\delta\phi$ respectively.
For these initial experiments EOM$\phi$ was not implemented.

L$_1$, L$_2$, M1, M2, and EOM$\theta$ are located on a fixed
plate. A preliminary alignment of the optics on the fixed plate
was performed before installation on the BEC apparatus.  In
particular, M1 and M2 were first aligned using a penta-prism, and
then lenses L$_1$ and L$_2$ were inserted and aligned to minimize
deflections. The entire plate was mounted next to the BEC
apparatus and the input lattice beam, $\vec{k}_1$, was aligned to
pass through the center of L$_1$ and L$_2$. With this technique we
measured that we were able to initially align the beams so that
the intersection angle deviated from orthogonality by only  $\mid
\epsilon \mid = $7 mrad.

\begin{figure}
\includegraphics[scale=.65]{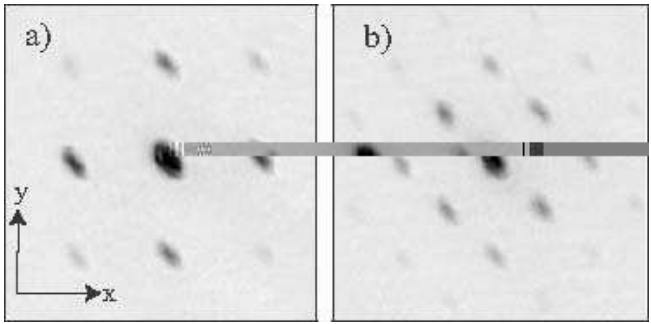}
\caption{Experimental images of atom diffraction from a 3 $\mu$s
pulse of a) the in-plane lattice lattice and b) the out-of-plane
lattice after 13 ms time-of-flight.}\label{diffraction}
\end{figure}

Calibration of the in-plane lattice depth is achieved by pulsing
the lattice and observing the resulting momentum distribution in
time-of-flight (TOF) \cite{Ovchinnikov1998}.  This atomic
diffraction pattern reveals the reciprocal lattice of the optical
lattice. Diffraction from the perfect in-plane lattice has
momentum components at multiples of $\pm 2 \hbar k \hat{x}$ and
$\pm 2 \hbar k \hat{y}$, while diffraction from the out-of-plane
lattice has additional components at multiples of $\pm
\sqrt{2}\hbar k (\hat{x} \pm \hat{y})$. The diffraction patterns
for both lattices after 13 ms TOF are shown in Fig.
\ref{diffraction}. For 120 mW and at $\lambda = 810$ nm, we
measure an average lattice depth of $U_0 = 40 $E$_R$ (E$_R =
\hbar^2 k^2/(2 m) = h \times 3.5$ kHz, m is the Rubidium mass) in
each of the independent 1D lattices making up the in-plane
lattice. As seen in Fig \ref{lattices}b, we calculate that the
out-of-plane lattice is four times deeper than the in-plane
lattice for equal intensity.

Pulsing the lattice is a useful method for determining the average
in-plane lattice depth, but this method discloses little information
about variations in depth $\Delta U$ between adjacent sites of the
in-plane lattice (such as variations caused by $\beta \neq 0$ and/or
$\epsilon \neq 0$). On the other hand, the ground state wave
function of the in-plane lattice is sensitive to $\Delta U$, and we
can use this to make $\beta, \epsilon \approx 0$. Information about
the ground state can be revealed by adiabatically loading the atoms
into the ground band of the lattice \cite{Denschlag2002}, quickly
switching off the lattice, and observing the atomic momentum
distribution in TOF. In this technique the lattice must be turned on
slowly enough to avoid vibrational excitation but quickly enough to
maintain phase coherence among sites; for our parameters the
timescale for loading is $\approx 500$ $\mu$s. (Note that band
adiabaticity is more complicated when we combine the in-plane and
out-of-plane lattices to create a double well lattice since the
tunnel couplings and tilt between double well sites can create
situations where band spacings are very small.) For a small but
nonzero $\Delta U$, this timescale is not adiabatic with respect to
tunneling between neighboring sites. In this way atoms are loaded
into every site, even though the true single particle ground state
fills every other site. Therefore, atoms are not in an eigenstate of
the potential, and the atomic wavefunction evolves in time.  In such
a lattice potential, pictured in Fig. \ref{imperfections}a, atoms in
adjacent sites acquire a differential phase, $ \Delta U t/\hbar$.
The ground band diffraction pattern changes in time as the atoms are
held in the lattice and the differential phase is allowed to ``wind
up".

\begin{figure}
\includegraphics[scale=.55]{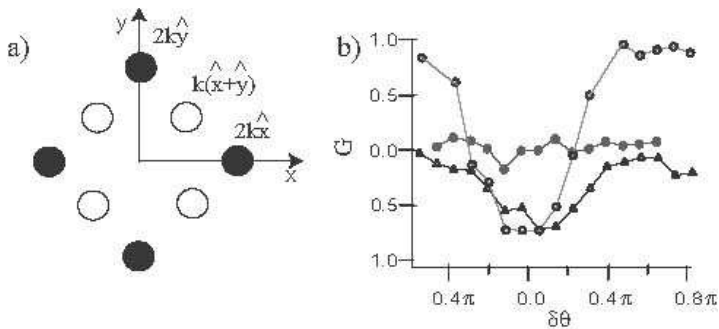}
\caption{a) Schematic of the momentum components that contribute to
$G = (N_{1 k } - N_{2k})/(N_{1 k} + N_{2k})$. $N_{2k}$ is the sum of
atoms in the momentum components designated with filled circles, and
$N_{1k}$ is the sum of atoms in momentum comonents designated with
open circles. b) Calibration of EOM$\beta$: $\beta \approx$ 52 mrad
(open circles), $\beta \approx$ 34 mrad (triangles), and $\beta
\approx$ 0 mrad (filled circles).}\label{beta}
\end{figure}

To quantify the ``ground band diffraction'' patterns, we define a
variable $G$ given by
\begin{equation}
   G =\frac{ N_{1 k } - N_{2k}}{N_{1 k} + N_{2k}}
\end{equation}
where  $N_{2k}$ is the number of atoms with momentum components $\pm
2 \hbar k \hat{x}$ and $\pm 2 \hbar k \hat{y}$,  and $N_{1k}$ is the
number of atoms with momentum components $\pm \sqrt{2}\hbar k
(\hat{x}\pm \hat{y})$ (see Fig. \ref{beta}a).   $G$ is normalized so
that the value $G= -1$ corresponds to a diffraction pattern
containing only momentum components associated with the in-plane
lattice.

We use the ground band diffraction to set the input polarization to
$\beta = 0$ by observing the dependence of the diffraction pattern
on the differential phase shift $\delta\theta$ at a fixed time. For
$\beta=0$ the light has no out-of-plane component so that changing
$\delta\theta$ with EOM$\theta$ does not change the topology but
merely translates the lattice. The calibration of EOM$\beta$ is done
by finding the condition in EOM$\beta$ which eliminates the effect
of EOM$\theta$, this corresponds to $\beta \approx 0$.  In practice
for a setting of EOM$\beta$, several ground band diffraction images
are analyzed at different values of $\delta\theta$. EOM$\beta$ is
then adjusted until scans of $\delta\theta$ produce no noticeable
difference in the diffraction pattern.

Sample data for the calibration of $\beta$ is shown in Fig.
\ref{beta}b. This method for determining $\beta = 0$ is convenient
because it is independent of other lattice imperfections, in
particular this method does not rely on $\epsilon = 0$.  For example
the optimal $\beta$ for the data shown in Fig. \ref{beta}b occurs
for $G \approx 0 \neq -1$.  $G \approx 0$ has no experimental
significance; it depends only on the time that the atoms were held
in the lattice. A perfect in-plane lattice would have $G = -1 $ for
all values of $\delta\theta$ at all values of time \footnote{Note
that $G = 1$ does not necessarily correspond to a perfect
out-of-plane lattice.}. The fact that $G \neq -1$ for $\beta = 0$
indicates the presence of momentum components at $\pm \sqrt{2}\hbar
k (\hat{x}\pm \hat{y})$ due to $\epsilon \neq 0$. With this method
we can set $\beta=0$ to zero within 17 mrad, placing an upper limit
on $\Delta U/U_0 \simeq 0.1\%$.

\begin{figure}
\includegraphics[scale=.80]{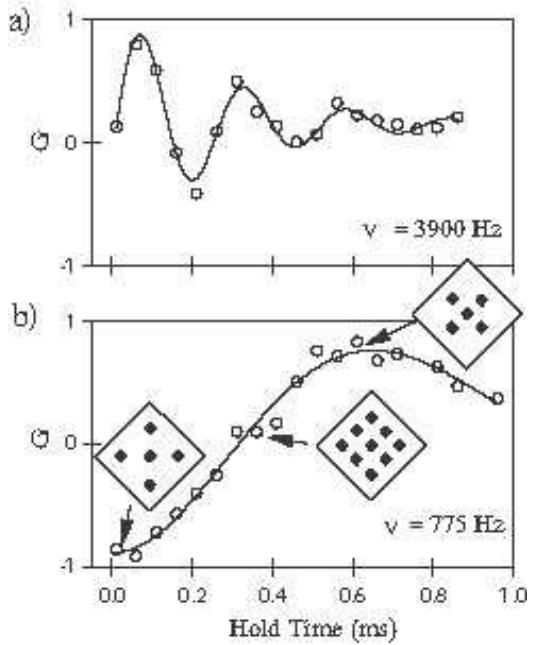}
\caption{Time dependence of the value $G$ characterizing the
diffraction patterns for atoms loaded into lattices with a small
offset energy $\Delta U$ caused by $\epsilon \neq 0$.  Open
circles are data points; solid lines are a fit to the data using a
exponentially decaying sinusoid. The frequency of the oscillations
given from the fit is inset in each image. From this frequency we
determine $\epsilon$:  a) $\nu = 3900$ Hz corresponds to $\epsilon
\simeq 7$ mrad, and b) $\nu = 775$ Hz corresponds to $\epsilon
\simeq 1.4$ mrad. The data in a) was taken after the initial
penta-prism alignment; the data in b) was taken after several
iterations of measuring the frequency and then realigning the
beams to further improve the angle. Schematics of the diffraction
patterns corresponding to different values of $G$ at different
times are shown in the insets. The initial phase of G in a) and b)
is arbitrary; it depends only on how much phase has been wound up
during the loading time. } \label{phase_osc}
\end{figure}

After setting $\beta = 0$ we determine $\epsilon$ by looking at the
time dependence of the ground band diffraction pattern.  We
adiabatically load the lattice in the method described above, then
we observe the time oscillations in the ground band diffraction
pattern varying between a diffraction pattern with $G=-1$ to $G=+1$.
From the time evolution of $G$ (see Fig. \ref{phase_osc}), we
extract the misalignment of the intersection angle, $\epsilon =
\Delta U/4U_0$. The data (open circles) in Fig. \ref{phase_osc} were
fit to an exponentially decaying sinusoid (solid line).

It is interesting to note the substantial decay in the amplitude of
the oscillations in $G$ shown in Fig. \ref{phase_osc}a, and the
reduced rate of decay in Fig. \ref{phase_osc}b.  We do not fully
understand this damping, or the reason why the damping is much less
for the improved $\epsilon$. Inhomogeneities in the lattice depth
due to the Gaussian nature of the lattice beams are not large enough
to account for the decay. However, factors such as mean field
effects, tunneling, and misalignments between the lattice beams and
the magnetostatic trap could contribute to the damping. Regardless
of the cause of the decay, we can use this method and the data shown
in Fig. \ref{phase_osc} to calculate and improve $\epsilon$.

From the fit to the time evolution of G we extract an oscillation
frequency, which can be used to calculate $\epsilon$. We calculate
$|\epsilon|$ after the initial penta-prism alignment to be 7 mrad
$\pm$ 0.2 mrad (Fig. \ref{phase_osc}a); the energy difference
between neighboring sites of a 40 E$_r$ lattice was 3.9 kHz $\pm$
100 Hz. We reduced $\epsilon$ by adjusting M2 and the optical flat
in order to change the angle of $\vec{k}_2$ while keeping the beam
aligned on the BEC, then remeasured the oscillation frequency of
$G$.   After several iterations of realignment and measurement, we
improved the alignment to $|\epsilon|=$1.4 mrad $\pm$ 0.2 mrad,
which corresponds to an energy offset of 775 Hz $\pm$ 70 Hz for a 40
E$_r$ lattice (Fig. \ref{phase_osc}b). For a 10 E$_r$ lattice the
energy offset would be $\lesssim 200$ Hz.  It is clear given the
signal-to-noise ratio in Fig. \ref{phase_osc}b that, if required,
the angle could be further improved.

We estimate the amount of polarization lattice from the measured
intensity imbalance of the four beams. The losses are due to
imperfect anti-reflective coatings on optical elements and the
uncoated  glass cell. The relative depth of the polarization lattice
is a function of $\alpha_v |\vec{E}^* \times \vec{E}|/\left(\alpha_s
|\vec{E}|^2\right)$ . For far-off-resonant traps the ratio
$(\alpha_v/\alpha_s)$ becomes small for the $5s_{1/2}$ ground state
of $^{87}$Rb \cite{Deutsch1998, Bonin}, thus decreasing the
polarization lattice depth. From Eq.~\ref{eq:Bfield}, the size of
the vector potential $U_{v}$ depends on the size and orientation of
the bias field. For our measured intensities, $I_2 = 0.85 I_1, I_3 =
0.81 I_1$,and $I_4 = 0.70 I_1$ with $\lambda = 810$ nm, we estimate
for the purely in-plane lattice that the maximum ratio $U_{v}/U_s$
in the barrier is $ \approx 8\%$.  In a combined in-plane and
out-of-plane lattice, the vector shift can lead to a state dependent
tilt.
%\footnote{To actually
%calculate the depth of the polarization lattice, one needs to
%account for the fact that the fictitious field created by the
%polarization lattice adds vectorially with other external magnetic
%fields. Depending on the size of those fields, we can make the
%ratio of $U_{v}/U_s$ as small as $\approx 3\%$.}.
For present experiments the vector shifts are not important, but in future experiments this could be useful to produce state dependent tunnel couplings and state dependent motion.

\section{Visualizing the Brillouin Zone}

\begin{figure}
\includegraphics[scale=.6]{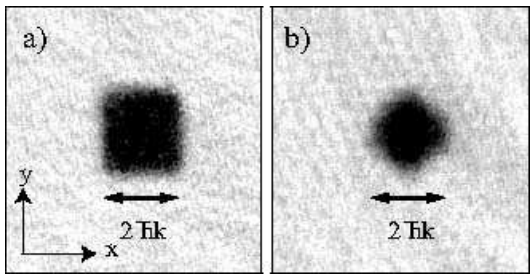}
\caption{Experimental images after 13 ms TOF of atoms filling the
first Brillouin zone for a) the in-plane lattice and b) the
out-of-plane lattice.  The shapes of the BZs reflect the momentum
components in each lattice.}\label{band_map}
\end{figure}

After minimizing the imperfections in the lattices, we can look at
the Brillouin zones (BZ) for each of the two lattices.  We load atoms into the      %site Casberg, Broweys
lattice in 100 ms, a timescale that is slow with respect to both
vibrational excitations and atom-atom interaction energies so that
atoms homogeneously fill the lowest band.  The lattice is then
turned off in 500 $\mu$s, mapping the atoms' quasi-momentum onto
free particle momentum states \cite{Kastberg1995,
Denschlag2002,Kohl2005,Greiner2001,Browaeys2005}. Atoms that
occupied the lowest energy band of a lattice will have momentum
contained in the first BZ of that lattice. The mapped zones for both
the in-plane and out-of-plane lattice are shown in Fig.
~\ref{band_map}.  As expected the bands are different for the
different lattices.  This is clear evidence that we have two
distinct lattices with distinct momentum components.

%By imaging the Brillouin zone of the in-plane and out-of-plane
%lattices, we have experimentally verified that the atoms are
%successfully transferred into bands that correspond to no
%vibrational excitation in the individual wells of the lattice
%\cite{Denschlag2002,Kohl2005,Greiner2001}.  We load atoms into the      %site Casberg, Broweys
%lattice in 100 ms, a timescale that is slow with respect to both
%vibrational excitations and interaction energies so that atoms
%homogeneously fill the lowest band.  The lattice is then turned
%off in 500 $\mu$s, mapping the atoms' quasi-momentum onto free
%particle momentum states. Atoms that occupied the lowest energy
%band of a lattice will have momentum contained in the first BZ of
%that lattice. The mapped zones for both the in-plane and
%out-of-plane lattice are shown in Fig. ~\ref{band_map}.

\section{ Dynamic Control of the double well lattice}

\begin{figure}
\includegraphics[scale=.85]{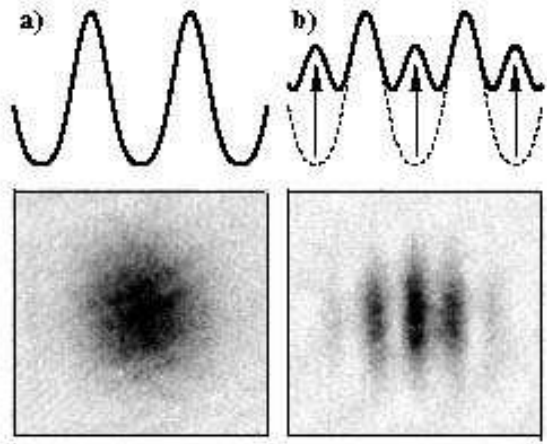}
\caption{a) Single-slit diffraction pattern resulting from a loss
of phase coherence among out-of-plane lattice sites.   b)
Double-slit interference pattern caused by coherence between atoms
within a particular double well but not among the ensemble of
double wells. }\label{fringes}
\end{figure}

As an example of the dynamic control of the double well lattice,
we demonstrate coherent splitting of atoms from single wells into
double wells. Initially, we load into the ground band of the
out-of-plane lattice.  The time scale for loading (100 ms) is
sufficiently slow to ensure dephasing of atoms in neighboring
sites.  If at this point in time we suddenly turn off the lattice
and allow 13 ms TOF, we observe a single, broad momentum
distribution, shown in Fig. \ref{fringes}a.  Since atoms on
separate sites have random relative phases, this distribution is
an incoherent sum of ``single-slit'' diffraction patterns from
each of the localized ground state wavefunctions in the
out-of-plane lattice.  The width of the single-slit pattern is
inversely proportional to the Gaussian width of the ground state
wave function in each lattice site \cite{Mandel2003}.

To demonstrate the coherent splitting of atoms, we start with the
ground-loaded out-of-plane lattice, then dynamically raise the
barrier to transfer the atoms into the symmetric double well
lattice.  The barrier is raised in 200 $\mu$s by increasing the
ratio of $I_{xy}/I_{z}$ with EOM$\beta$, while EOM$\theta$ is set
to $\delta \theta = \pi/2$ \footnote{We experimentally set $\delta
\theta = \pi/2$ by observing the double slit diffraction pattern
(Fig. \ref{fringes}b) and adjusting $\delta\theta$ until the
diffraction pattern is symmetric.  For $\delta \theta \neq \pi/2$
the diffraction pattern would be shifted to the right or left.}.
This timescale is chosen to be slow enough to avoid vibrational
excitations but fast enough to maintain phase coherence within a
double well. Since there is no phase coherence from one double
well to another, the resulting momentum distribution is an
incoherent sum of essentially identical double-slit diffraction
patterns (shown in Fig. \ref{fringes}b) from each of the
wavefunctions localized in individual double wells.

\section{ Conclusions}

The ability to isolate individual atoms in controllable double
well potentials is essential for testing a variety of neutral atom
based quantum gate proposals.  Two-qubit gate ideas typically
involve state dependent motion \cite{Brennen1999, Jaksch1999} or
controlled state dependent interaction \cite{Calarco2004}, but
nearly all require the ability to move atoms into very near
proximity (e.g. into the same site) and subsequently to separate
them.  The flexibility and dynamic control of the double well
lattice can be used to demonstrate and test motion of atoms
between wells.  Furthermore, state dependence of the barrier
height can be used for state dependent motion between wells, allowing for the possibility of 2-atom gates.

In conclusion we have demonstrated a dynamically controllable
double-well lattice. %suitable for implementing and testing two
%qubit gates.
The geometry of this lattice is topologically phase stable against
vibrational noise, yet allows topological control of the lattice
structure. The design of the double well lattice allows for
flexible real-time control of its properties:  the tilt and the
tunnel barrier between sites within the double well. In addition,
the orientation of the double well can be adjusted so that a site
can be paired with any one of its four nearest neighbors. We have
described technical issues and imperfections of the double well
lattice, and we have presented techniques to minimize the
imperfections. We have demonstrated dynamic control of the double
well lattice by showing the coherent transfer of atoms from single
wells to double wells. In the future, the double well lattice
presented here could be used for applications in quantum
computation and quantum information processing, as well as
studying interesting extensions of the Bose-Hubbard model, such as
the emergence of super-solids and density waves
\cite{Scarola2005}.

\begin{acknowledgments}
The authors would like to acknowledge William D. Phillips for a
critical reading of this manuscript, and WDP and Steve Rolston for
many enlightening and insightful conversations. This work was
supported by ARDA, ONR, and NASA. J.S-S acknowledges support from
the NRC postdoctoral research program.  PSJ acknowledges support
from NSF.
\end{acknowledgments}

\newcommand{\noopsort}[1]{} \newcommand{\printfirst}[2]{#1}
   \newcommand{\singleletter}[1]{#1} \newcommand{\switchargs}[2]{#2#1}

\end{document}